\documentclass[journal]{IEEEtran}

\usepackage{wrapfig}
\usepackage{booktabs}
\usepackage{algorithm}
\usepackage{algorithmic}
\usepackage{amsmath}
\usepackage{xcolor}
\usepackage{caption}
\usepackage{pifont}
\usepackage{multirow}
\usepackage[colorlinks, citecolor=blue]{hyperref}
\definecolor{deepblue}{rgb}{0, 0, 0.9}
\definecolor{lightgray}{rgb}{0.4, 0.4, 0.4}
\usepackage{arydshln}
\usepackage{graphicx}
\usepackage{wrapfig}
\usepackage{xcolor}         
\usepackage{hyperref}
\usepackage{subcaption}
\usepackage{url}            
\usepackage{amsfonts}       
\usepackage{nicefrac}       
\usepackage{microtype}      
\usepackage{tikz}

\usepackage{textpos}

\usepackage{eso-pic}
\usepackage{atbegshi}
\usepackage{picture}

\begin{document}
\title{Distinctive Feature Codec: An Adaptive Efficient Speech Representation for Depression Detection}



\author{Xiangyu Zhang~\IEEEmembership{Student Member,~IEEE,}, Fuming Fang, Peng Gao, Bin Qin, Beena Ahmed~\IEEEmembership{Member,~IEEE,}, Julien Epps~\IEEEmembership{Senior Member,~IEEE}
\thanks{Xiangyu Zhang, Beena Ahmed, Julien Epps are with the School of Electrical Engineering and Telecommunications, University of New South Wales, Sydney, Australia.}
\thanks{Fuming Fang, Peng Gao, Bin Qin are with Xiaomi Corp, Beijing, China.}
}


\maketitle

\begin{abstract}
Large Language Models (LLMs) have demonstrated remarkable success across diverse fields, establishing a powerful paradigm for complex information processing. This has inspired the integration of speech into LLM frameworks, often by tokenizing continuous audio via neural speech codecs, enabling powerful speech language models. However, this dominant tokenization strategy relies on uniform frame-based processing at fixed time intervals. This fixed-rate approach, while effective for linguistic content, destroys the temporal dynamics. These dynamics are not noise but are established as primary biomarkers in clinical applications such as depression detection. To address this gap, we introduce the Distinctive Feature Codec (DFC), an adaptive framework engineered to preserve this vital timing information. Drawing from linguistic theory, DFC abandons fixed-interval processing and instead learns to dynamically segment the signal at perceptually significant acoustic transitions. This generates variable-length tokens that efficiently encode the temporal structure. As a key contribution, this work is the first to integrate traditional distinctive features into a modern deep learning codec for a temporally sensitive task such as depression detection. We also introduce the Group-wise Scalar Quantization (GSQ) approach to stably quantize these variable-length segments. Our distinctive feature-based approach offers a promising alternative to conventional frame-based processing and advances interpretable representation learning in the modern deep learning speech depression detection framework.
\end{abstract}


\section{Introduction}

The remarkable success of large language models (LLMs) in understanding and generating text~\cite{bai2023qwen,dubey2024llama,touvron2023llama} has inspired researchers to develop similar architectures for speech processing~\cite{defossez2024moshi,du2024cosyvoice,du2024cosyvoice2}. This expansion is motivated by the rich information encoded in speech signals beyond mere linguistic content, including speaker identity, emotion, and prosody~\cite{borsos2023audiolm,nguyen2023generative,lakhotia2021generative,zhang2023speechgpt}. A fundamental challenge in building these speech-aware models is the tokenization of continuous audio into representations that neural networks can process. Unlike text, which has natural boundaries~\cite{yu2023megabyte,the2024large,pagnoni2024byte}, speech is continuous and complex. Consequently, most current approaches—whether for discrete tokenization~\cite{defossezhigh,zeghidour2021soundstream,zhang2024speechtokenizer} or self-supervised feature extraction~\cite{baevski2020wav2vec,hsu2021hubert}—primarily rely on \textbf{frame-based processing} with fixed time intervals.

While this fixed-rate processing is effective for tasks focused on linguistic content, such as speech recognition, it presents a fundamental limitation: it overlooks the varying information density of speech and, critically, \textbf{destroys the fine-grained temporal dynamics}~\cite{stevens2002toward}. This frame-based segmentation arbitrarily cuts through natural speech events, disrupting prosodic rhythms, distorting pause structures, and obscuring speech rate variations. These temporal dynamics are not noise; they are established as primary biomarkers in critical clinical applications, most notably depression detection~\cite{huang2019investigation,zhang2025speecht}. Thus, a fundamental mismatch exists: the dominant speech representation methods are optimized for linguistic content, making them inherently unsuitable for downstream tasks that depend on temporal fidelity.


\begin{figure}[t]
    \centering
    \includegraphics[width=0.45\textwidth]{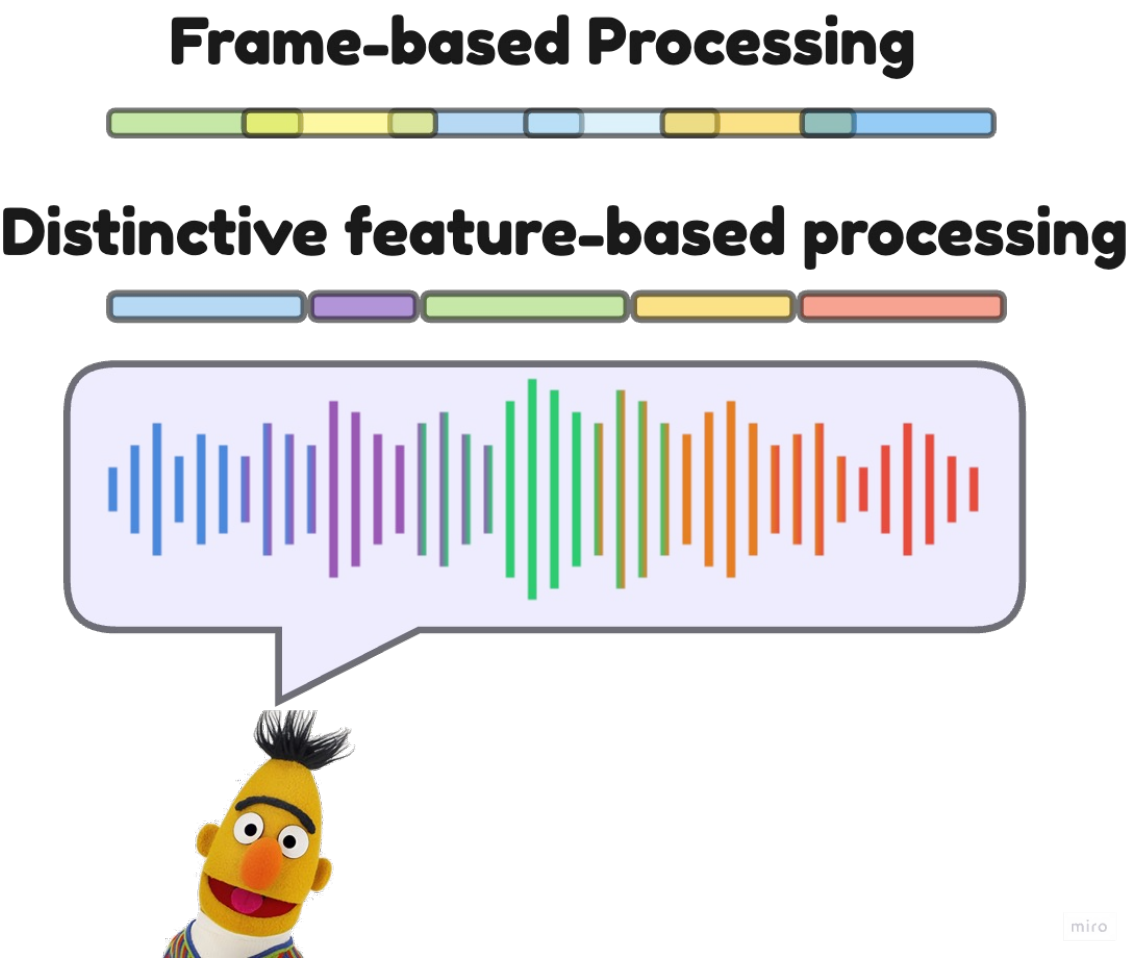}
    \caption{Comparison of segmentation strategies. The Frame-based processing (top) imposes a rigid grid of fixed-length intervals, often disrupting natural speech events. In contrast, Distinctive feature-based processing (bottom) adaptively segments the signal by identifying perceptually significant acoustic transitions as boundaries. This results in variable-length segments that preserve the fine-grained temporal dynamics essential for clinical tasks}
    \label{fig:motivation}
\end{figure}
To bridge this gap, we must identify meaningful units within continuous signals in a way that preserves temporal structure. The linguistic theory of distinctive features provides an insightful alternative~\cite{jakobson1952preliminaries}. This theory posits that speech can be naturally segmented at points where acoustic characteristics are most differentiated~\cite{liu1996landmark,zhang2024auto}. Instead of arbitrary fixed-length segments, this approach identifies boundaries where acoustic properties undergo significant changes. Such an adaptive, variable-length segmentation process naturally encodes the temporal dynamics; for example, a pause is no longer a sequence of identical "silence" frames, but is represented by a single, long segment whose duration is explicitly preserved. However, implementing distinctive feature analysis in neural networks has remained challenging. The irregular, variable-length nature of these features conflicts with the regular, grid-like computations (e.g., convolutions, transformers) that dominate modern deep learning~\cite{lecun2015deep,krizhevsky2012imagenet, baevski2020wav2vec,hsu2021hubert}. This technical mismatch has largely confined distinctive feature analysis to traditional signal processing approaches~\cite{stevens2002toward,li2008contribution}.

In this paper, we address this challenge by introducing the \textbf{Distinctive Feature Codec (DFC)}, a framework designed to learn an efficient speech representation that preserves the critical temporal information required for depression detection. We propose the first architecture that successfully integrates the theory of distinctive features into a modern, end-to-end neural codec framework~\cite{sutskever2014sequence,hinton2006reducing}. Our approach uses a lightweight, self-supervised boundary detector to identify perceptually significant acoustic transitions, which guide an adaptive encoder-decoder (based on SEANet~\cite{tagliasacchi2020seanet}) to process variable-length segments. This marks a departure from conventional fixed-interval processing~\cite{defossezhigh,zeghidour2021soundstream,zhang2024speechtokenizer}. Furthermore, our investigation reveals that standard quantization methods like Finite Scalar Quantization (FSQ)~\cite{mentzerfinite} become unstable when applied to such variable-length segments at low bitrates. To solve this, we develop a novel \textbf{Group-wise Scalar Quantization (GSQ)} approach, which ensures robust and stable quantization. Our work validates distinctive features as a promising direction for codec design, offering new perspectives on efficient speech representation for depression detection.

\section{Preliminary}
\subsection{Distinctive Features} 
Distinctive features, first proposed in linguistic theory~\cite{jakobson1952preliminaries,liu1996landmark,lee1988automatic,zhang2024auto}, characterize speech by identifying regions with acoustically distinctive properties that help differentiate speech segments from one another. As illustrated in Fig.\ref{fig:motivation}, this approach fundamentally differs from conventional frame-based processing: while frame-based methods uniformly segment speech signals into fixed-length overlapping windows, distinctive feature analysis identifies boundaries where acoustic characteristics undergo significant changes. This approach has proven valuable in early automatic speech recognition systems~\cite{liu1996landmark,lee1988automatic,he2019ctc} and has been successfully applied to various healthcare applications~\cite{ishikawa2017toward,ishikawa2023landmark,zhang2024llms,zhang2025speecht,zhang2025pre}. However, despite its theoretical advantages, the development of distinctive feature-based methods has faced significant limitations in the deep learning era. Traditional implementations of distinctive features heavily rely on linguistic expertise and hand-crafted rules~\cite{liu1996landmark,zhang2024auto}, leading to limited training data and difficulties in scaling across different acoustic conditions and languages. Additionally, while frame-level processing naturally aligns with convolutional neural networks and enables efficient batch processing, the variable-length nature of distinctive features poses challenges for modern deep learning architectures. The success of frame-level processing in various deep learning systems~\cite{baevski2020wav2vec,hsu2021hubert} has led to a rich ecosystem of pre-trained models and established practices, making it the predominant choice for modern speech processing systems despite its inherent inefficiencies. This technical mismatch, combined with the limited availability of labeled data for distinctive feature analysis, has constrained its adoption in contemporary deep learning approaches.

\subsection{Speech Codecs} 
Speech codecs compress speech signals into discrete tokens while preserving essential acoustic and linguistic information~\cite{vinton2001scalable,epps2000wideband}. These discrete representations serve as inputs for downstream tasks such as large language models~\cite{du2024cosyvoice,du2024cosyvoice2,defossez2024moshi}. Current approaches can be broadly categorized into two paradigms: \textbf{end-to-end trained codecs} that directly learn discrete representations through reconstruction objectives~\cite{zeghidour2021soundstream,defossezhigh,zhang2024speechtokenizer}, and \textbf{two-stage approaches} that first extract semantic features using self-supervised models~\cite{hsu2021hubert,baevski2020wav2vec}, then apply generative modeling techniques such as flow matching~\cite{lipmanflow,mehta2024matcha} or diffusion~\cite{ho2020denoising,zhang2023speak}.

Two-stage approaches (e.g., Sylber~\cite{cho2024sylber}) achieve impressive compression by operating at syllable-level granularity~\cite{bpe,baade2024syllablelm}, reducing token rates to as low as 5-10 tokens per second. These methods prioritize \textit{semantic preservation}—retaining linguistic content that overlaps with text representations—while relying on powerful generative models. However, this design philosophy is fundamentally misaligned with depression detection, where diagnostic information resides not in semantic content but in \textit{acoustic characteristics}: voice quality, spectral energy distribution, and prosodic micro-variations~\cite{zhang2025speecht}.

Table~\ref{tab:sylber_codec},~based on our experimental evaluation of the Sylber model, empirically demonstrates this limitation. As Sylber's model capacity increases (K-means clusters from 10K to 20K), perceptual quality (PESQ) paradoxically \textit{degrades} from 1.25 to 0.81, while mel-spectral error increases from 0.616 to 0.673. This pattern indicates that the generative model produces perceptually plausible speech by hallucinating acoustic details rather than preserving them from the original signal. While the semantic content ("what was said") remains intact, the acoustic substrate ("how it was said") diverges from the input—erasing the subtle spectral and prosodic deviations that constitute depression biomarkers. Multiple studies confirm that such semantically-focused representations underperform on tasks requiring acoustic fidelity~\cite{cuervo2024scaling,wang2024speech}.

\begin{table}[t]
\caption{Analysis of Sylber Codec performance. The table shows mel\_error, stft, pseq, and stoi metrics for different KM (K-Means) cluster sizes.}
\label{tab:sylber_codec}
\centering
\begin{tabular}{lcccc}
\toprule
\textbf{Sylber Codec} & \textbf{mel\_error} $\downarrow$ & \textbf{stft} $\downarrow$ & \textbf{PESQ} $\uparrow$ & \textbf{stoi} $\uparrow$ \\
\midrule
KM=5K  & 0.6204 & 2.0246 & 1.0025 & 0.7066 \\
KM=10K & 0.6155 & 2.2049 & 1.2501 & 0.7136 \\
KM=20K & 0.6727 & 2.1525 & 0.8135 & 0.7003 \\
\bottomrule
\end{tabular}
\end{table}
\begin{figure*}[t]
  \centering
  \includegraphics[width=0.8\textwidth]{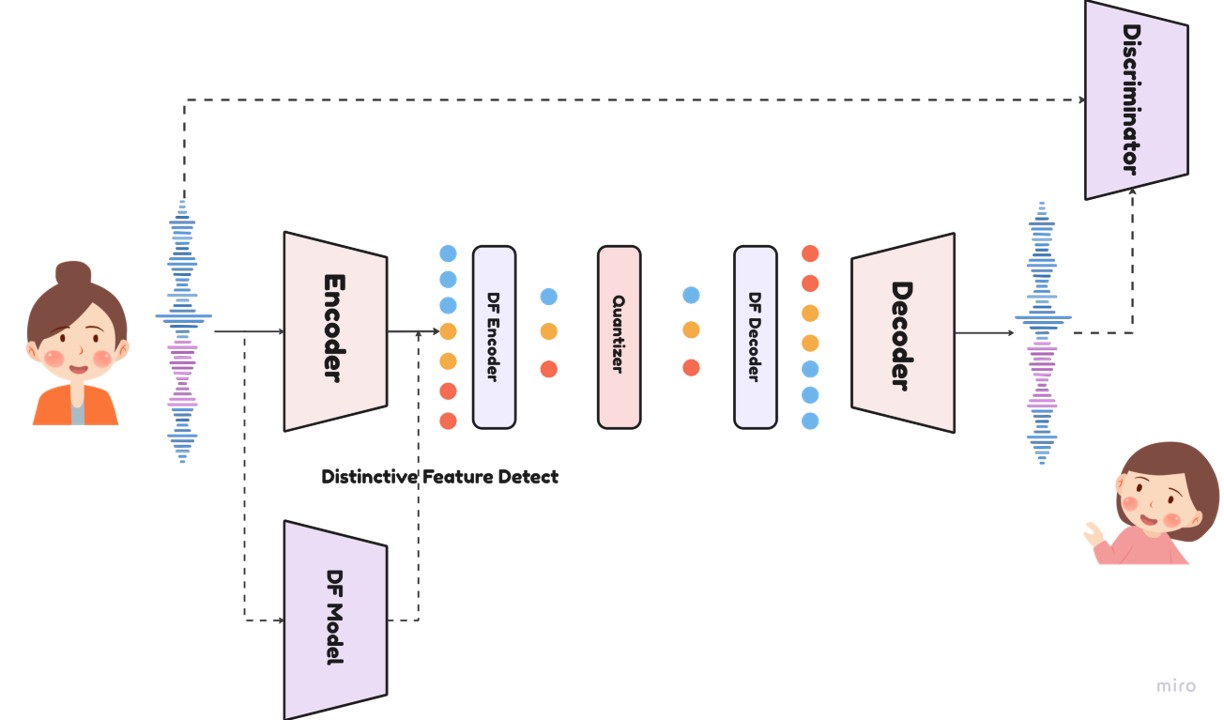}
  \caption{Overview of the Distinctive Feature Codec framework. The Distinctive Feature Detector (top) identifies acoustic boundaries through contrastive learning to guide variable-length segmentation. The codec pipeline segments encoded speech based on these boundaries, compresses each segment via the DF Encoder, applies quantization, and reconstructs through the DF Decoder and main decoder.}
  \label{fig:overview}
\end{figure*}

Our work therefore focuses on \textbf{end-to-end trained codecs}, which optimize for faithful acoustic reconstruction rather than semantic compression, explicitly preserving the spectral and temporal dynamics essential for clinical applications.

\subsection{Speech Representation for Depression Detection}

Clinical research has long established that depression profoundly affects speech production mechanisms, manifesting in distinctive acoustic biomarkers known as ``psychomotor retardation''~\cite{sobin1997psychomotor}. 
These manifestations primarily include reduced speech rate, monotonic prosody, prolonged pauses, and blunted articulation~\cite{mundt2012vocal, quatieri2012vocal}. 
Consequently, capturing these \textbf{temporal dynamics} and \textbf{prosodic variations} has been a central focus in feature extraction for depression detection.

Early approaches predominantly relied on hand-crafted Low-Level Descriptors (LLDs) to explicitly model these characteristics. Standard feature sets, such as eGeMAPS and INTERSPEECH ComParE~\cite{eyben2015geneva},
aggregate frame-level acoustic properties (e.g., pitch, energy, formants) using statistical functionals to capture global prosodic trends.
More distinctively, some studies explored \textbf{landmark-based} or \textbf{distinctive feature} analysis~\cite{huang2019investigation,zhang2025pre},
which focuses on specific acoustic events---such as the onset of bursts or glottal transitions---to detect subtle articulatory coordination deficits associated with depression~\cite{huang2019investigation, ishikawa2017toward, zhang2025speecht}.
These methods offered high interpretability and preserved the temporal integrity of speech events but were limited by their inability to model complex, high-level abstractions.

In the deep learning era, the paradigm shifted towards learning data-driven representations directly from raw audio or spectrograms using Convolutional Neural Networks (CNNs) and Transformers~\cite{lecun2015deep, krizhevsky2012imagenet}.
Self-supervised models (SSL) like Wav2Vec 2.0 and HuBERT have achieved state-of-the-art performance by encoding speech into continuous or discrete representations~\cite{baevski2020wav2vec, hsu2021hubert, zeghidour2021soundstream, zhang2024speechtokenizer}.
However, these modern approaches predominantly employ \textbf{fixed-rate frame-based processing}, where speech is segmented into uniform time intervals regardless of the underlying acoustic content. While effective for linguistic tasks like ASR, recent studies suggest that this rigid segmentation may disrupt the fine-grained temporal structures---such as variable-length pauses and rhythmic patterns---that serve as critical diagnostic cues~\cite{stevens2002toward, zhang2025pre, zhang2025speecht}.
This creates a fundamental mismatch: the dominant feature extraction methods are optimized for semantic continuity, potentially at the cost of the temporal fidelity required for reliable depression assessment.

\section{Distinctive Codec Framework}
As illustrated in Fig.~\ref{fig:overview}, our approach consists of two key stages: First, a lightweight boundary detector is trained to identify perceptually significant transitions in speech signals. Second, these detected boundaries guide the codec model to adaptively merge or separate speech segments, leading to more efficient tokenization that aligns with the natural structure of speech.
\subsection{Distinctive Features Detector}\label{Distinctive Features Detector}
The core idea of distinctive features lies in identifying regions where speech segments exhibit maximal acoustic contrast with their neighbors. This naturally aligns with the objective of contrastive learning, which aims to learn representations by maximizing the differences between distinct samples while minimizing differences between similar ones~\cite{khosla2020supervised,gao2021simcse,reimers2019sentence,kreuk2020self}. We leverage this connection to design a self-supervised boundary detector that learns to identify distinctive features without requiring phoneme-level annotations.

Specifically, we train a lightweight encoder network $f_\theta: \mathbb{R}^L \rightarrow \mathbb{R}^D$ that maps raw speech input segments $\mathbf{x} = \{x_1,...,x_T\}$ into a latent representation space, where $L$ is the segment length and $D$ is the latent dimension. For a given segment $x_t$ at position $t$, we compute its latent representation $\mathbf{z}_t = f_\theta(x_t)$ and compare it with subsequent segments at different positions ${t+k}_{k=1}^K$. The similarity score is computed along the feature dimension:

\begin{equation}
    s(t,k) = - \alpha \cdot \text{cos}(\mathbf{z}_t, \mathbf{z}_{t+k})_D
\end{equation}

where $\alpha$ is a scaling coefficient. For each positive pair $(x_t, x_{t+k})$, we construct a set of negative samples by randomly shuffling segments from the same batch. The model is trained to minimize the contrastive loss:

\begin{equation}
\mathcal{L} = -\mathbb{E}_{t,k}[\log \frac{\exp(s(t,k)/\tau)}{\exp(s(t,k)/\tau) + \sum_{n=1}^N \exp(s(t,n)/\tau)}]
\end{equation}

where $\tau$ is a temperature parameter and $N$ is the number of negative samples. This contrastive objective encourages the model to learn representations that capture the inherent acoustic differences between speech segments. The resulting similarity scores naturally highlight regions where acoustic characteristics undergo significant changes, corresponding to distinctive feature boundaries. These boundaries then guide the subsequent merging of segments in our codec model.

\subsection{Distinctive Codec}
To evaluate our distinctive feature-based approach, we build our codec model on top of the SpeechTokenizer framework~\cite{zhang2024speechtokenizer}, which leverages the SEANet architecture~\cite{tagliasacchi2020seanet} for encoder-decoder operations. The key innovation of our approach lies in how we process the encoded representations based on the distinctive features detected by our boundary detector.
Given input speech $\mathbf{x} \in \mathbb{R}^{1 \times L}$, the encoder $E_{\theta}$ first maps it to a latent representation:
\begin{equation}
\mathbf{e} = E_{\theta}(\mathbf{x}) \in \mathbb{R}^{D \times T}
\end{equation}
where $D$ is the feature dimension and $T = L/r$ represents the temporal dimension after downsampling with ratio $r$.
Traditional frame-based codecs would typically process this representation uniformly across time. In contrast, our distinctive codec first identifies segment boundaries $\mathcal{B} = {b_1, b_2, \ldots, b_M}$ using the boundary detector described in Section~\ref{Distinctive Features Detector}. With these boundaries, we partition the feature sequence into variable-length segments:
\begin{equation}
\mathbf{S} = {\mathbf{s}_1, \mathbf{s}_2, \ldots, \mathbf{s}_{M+1}}
\end{equation}
where $\mathbf{s}_i = \mathbf{e}[:, b_{i-1}:b_i]$ represents the $i$-th segment ($b_0=0$ and $b_{M+1}=T$ for notational convenience).
For each segment, we apply a Distinctive Feature encoder (DFE) to compress the variable-length representation into a fixed-length embedding:
\begin{equation}
\mathbf{z}_i = \text{DFE}(\mathbf{s}_i) \in \mathbb{R}^{H \times 1}
\end{equation}
where $H$ is the hidden dimension. This operation effectively merges temporal information within each segment into a single token, guided by the distinctive feature boundaries.
During decoding, we expand each compressed segment embedding back to its original length using a Distinctive Feature decoder (DFD):
\begin{equation}
\hat{\mathbf{s}}_i = \text{DFD}(\mathbf{z}_i, l_i)
\end{equation}
where $l_i = b_i - b_{i-1}$ is the original segment length. The full sequence is reconstructed by concatenating the expanded segments:
\begin{equation}
\hat{\mathbf{e}} = \text{Concat}(\hat{\mathbf{s}}_1, \hat{\mathbf{s}}_2, \ldots, \hat{\mathbf{s}}_{M+1})
\end{equation}
Finally, the decoder $D_{\phi}$ transforms the reconstructed latent representation back to the waveform domain:
\begin{equation}
\hat{\mathbf{x}} = D_{\phi}(\hat{\mathbf{e}})
\end{equation}

\subsection{Group-wise Scalar Quantization}
Finite Scalar Quantization (FSQ)~\cite{mentzerfinite} has emerged as an effective approach for discrete representation learning due to its computational efficiency and strong performance across various tasks~\cite{du2024cosyvoice2,yu2024language,parker2024scaling}. Unlike vector quantizers that require nearest neighbor search in high-dimensional spaces, FSQ directly quantizes each dimension of the latent representation independently, significantly reducing computational complexity.


\begin{figure}[t]
    \centering
    \includegraphics[width=0.38\textwidth]{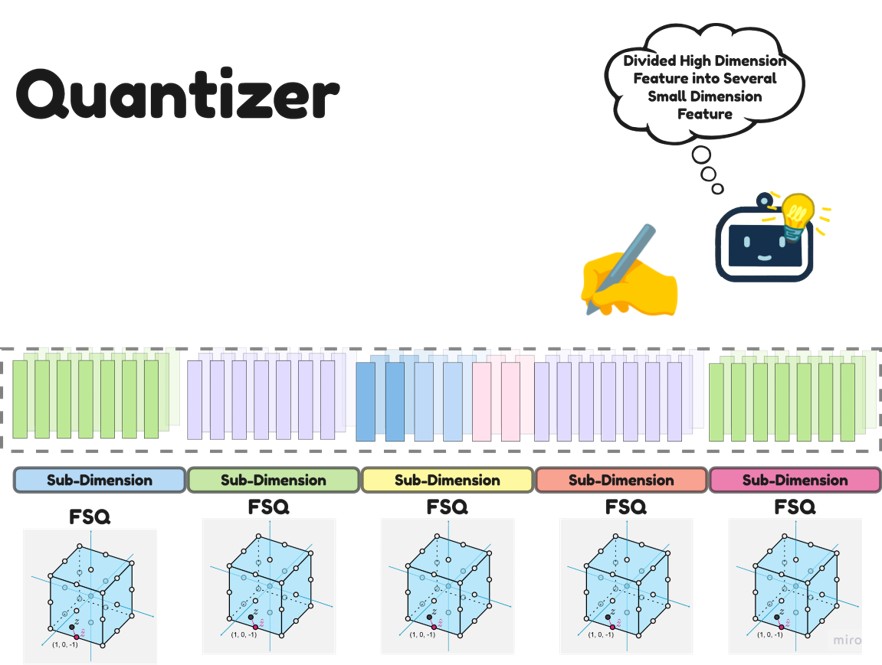}
    \caption{ Features are divided into smaller groups for independent quantization, enhancing stability and representation quality.}
    \label{fig:quantizer}
\end{figure}

However, during our experiments with distinctive feature-based tokenization, we discovered that standard FSQ becomes unstable when operating at high downsampling rates. This instability manifests as training divergence and poor reconstruction quality, particularly when compressing longer, variable-length segments into single tokens. We hypothesize that this issue stems from the increased difficulty of directly quantizing high-dimensional features with varying temporal characteristics.

To address this issue, we propose Group-wise Scalar Quantization (GSQ) as shown in Figure~\ref{fig:quantizer}, which decomposes the high-dimensional quantization problem into multiple lower-dimensional sub-problems. Given a compressed segment representation $\mathbf{z}_i \in \mathbb{R}^{H \times 1}$, we divide it into $G$ groups, each with dimension $H_g = H/G$:
\begin{equation}
\mathbf{z}_i = [\mathbf{z}_i^1, \mathbf{z}_i^2, \ldots, \mathbf{z}_i^G]
\end{equation}

Each group is then processed independently through quantization. The key design choice lies in how we parameterize the quantization function for each group $Q_g(\cdot)$. We explore two variants:

\noindent\textbf{Direct Quantization (M2M):} Each group is quantized directly without dimensional change:
\begin{equation}
\hat{\mathbf{z}}_i^g = \text{FSQ}_g(\mathbf{z}_i^g), \quad \mathbf{z}_i^g, \hat{\mathbf{z}}_i^g \in \mathbb{R}^{H_g}
\end{equation}
This preserves the distributed representation across all $H_g$ dimensions but requires a codebook of size $L^{H_g}$ per group (where $L$ is the number of quantization levels per dimension).

\noindent\textbf{Projection-based Quantization (M2O):} Each group is first projected to a lower-dimensional space (in our case, a scalar), quantized, then projected back:
\begin{equation}
p_i^g = \mathbf{W}_g^{(c)} \mathbf{z}_i^g, \quad \hat{p}_i^g = \text{FSQ}_g(p_i^g), \quad \hat{\mathbf{z}}_i^g = \mathbf{V}_g^{(e)} \hat{p}_i^g
\end{equation}
where $\mathbf{W}_g^{(c)} \in \mathbb{R}^{k \times H_g}$ is the compression matrix, $\mathbf{V}_g^{(e)} \in \mathbb{R}^{H_g \times k}$ is the expansion matrix, and $k$ is the projection dimension (we use $k=1$, i.e., scalar projection). This reduces the codebook size to $L^k$ per group while maintaining expressiveness through learned projections.

The full representation is reconstructed by concatenating all quantized groups:
\begin{equation}
\hat{\mathbf{z}}_i = [\hat{\mathbf{z}}_i^1, \hat{\mathbf{z}}_i^2, \ldots, \hat{\mathbf{z}}_i^G]
\end{equation}

This group-wise decomposition is particularly beneficial for quantizing variable-length segments with diverse acoustic characteristics. By processing features in smaller groups with dedicated quantizers, we reduce quantization complexity through lower-dimensional operations. Our primary experiments employ the M2O variant with scalar projection ($k=1$) due to its superior stability at low token rates and dramatically reduced codebook size. Algorithm~\ref{alg:gsq} presents the unified framework, with the projection step (lines 3-4, 6) being optional for the M2M variant.

\begin{algorithm}[t]
\caption{Group-wise Scalar Quantization (GSQ)}
\label{alg:gsq}
\begin{algorithmic}[1]
\REQUIRE Input tensor $x \in \mathbb{R}^{B \times T \times D}$
\REQUIRE Number of groups $G$, group size $d = D / G$
\REQUIRE Use projection: $\texttt{use\_proj} \in \{\text{true}, \text{false}\}$
\REQUIRE Projection dimension: $k$ (typically $k=1$ for M2O)
\REQUIRE \textit{If} $\texttt{use\_proj}$: compression matrices $\{W_i^{(c)} \in \mathbb{R}^{k \times d}\}_{i=1}^G$, expansion matrices $\{W_i^{(e)} \in \mathbb{R}^{d \times k}\}_{i=1}^G$
\REQUIRE FSQ quantizers $\{\text{FSQ}_i\}_{i=1}^G$
\ENSURE Quantized output $\hat{x} \in \mathbb{R}^{B \times T \times D}$
\STATE
\FOR{$i = 1$ \TO $G$}
    \STATE $x_i \leftarrow x[:, :, i \cdot d : (i+1) \cdot d]$ \COMMENT{Extract group $i$}
    \STATE \textbf{if} $\texttt{use\_proj}$ \textbf{then} \COMMENT{M2O: project to lower dim}
    \STATE \quad $x_i \leftarrow W_i^{(c)} x_i$ 
    \STATE \textbf{end if}
    \STATE $\hat{x}_i \leftarrow \text{FSQ}_i(x_i)$ \COMMENT{Quantize (M2M or M2O)}
    \STATE \textbf{if} $\texttt{use\_proj}$ \textbf{then} \COMMENT{M2O: project back}
    \STATE \quad $\hat{x}_i \leftarrow W_i^{(e)} \hat{x}_i$
    \STATE \textbf{end if}
\ENDFOR
\STATE
\STATE $\hat{x} \leftarrow \text{Concat}([\hat{x}_1, \hat{x}_2, \ldots, \hat{x}_G], \text{dim}=-1)$
\RETURN $\hat{x}$
\end{algorithmic}
\end{algorithm}

\section{Experimental Setup}
We implement our Distinctive Codec using the widely-adopted SEANet-based~\cite{tagliasacchi2020seanet} encoder-decoder architecture, which has become the de facto standard in modern neural speech codecs including SoundStream~\cite{zeghidour2021soundstream}, EnCodec~\cite{kumar2023high}, DAC~\cite{kumar2023high}, and SpeechTokenizer~\cite{zhang2024speechtokenizer}. This architectural choice enables direct comparison with existing methods while isolating the impact of our distinctive feature-based adaptive segmentation versus conventional fixed-rate processing. The encoder employs strided convolutional layers for downsampling, bidirectional LSTM layers for temporal modeling, and residual blocks for feature refinement. The decoder mirrors this structure to reconstruct the waveform from the latent representation.

For training and evaluation, we use the LibriSpeech dataset~\cite{panayotov2015librispeech}, a widely-adopted benchmark containing 960 hours of read English speech with diverse speakers and acoustic conditions. We train on the standard train set and evaluate codec reconstruction quality on 500 randomly selected samples from the test-clean set. This evaluation protocol maintains consistency with prior codec research and enables direct performance comparison across methods. For downstream evaluation of depression detection, we use the DAIC-WOZ dataset~\cite{devault2014simsensei}, which contains clinical interviews specifically designed for mental health assessment and provides naturalistic speech with diagnostically relevant temporal dynamics.
\subsection{Details of Distinctive Feature Detector}
The Distinctive Feature Detector, a core component of our framework, was implemented as a lightweight CNN-based architecture designed to identify perceptually significant transitions in speech signals using contrastive learning. 

\paragraph{Model Architecture} The detector uses a 5-layer CNN structure processing raw audio input directly. Each layer consists of a 1D convolutional operation followed by batch normalization and LeakyReLU activation. The network employs variable kernel sizes and strides to progressively downsample the input while capturing acoustic patterns at different time scales. For our primary configuration which  yields 9.5 tokens per second (as shown in Table \ref{tab:speech_tokenization}), the convolutional layers use kernel sizes of 10, 8, 8, 4, and 4, with corresponding stride values of 5, 4, 4, 2, and 2. For our higher frame rate configuration (15.7 tokens per second), we adjusted the stride values to 5, 4, 4, 4, and 2, demonstrating the adaptability of our approach. This flexibility allows our distinctive feature-based method to operate effectively across different token rates, unlike approaches such as Sylber Codec that are limited to specific syllable-level rates. Our configurable architecture provides an effective receptive field capable of capturing both local and broader acoustic transitions while allowing token rate adjustments based on application requirements.

The network's final embedding dimension was set to 256, with an optional projection layer that could further reduce this dimension to 64 for more compact representations. We found that applying this projection with a linear transformation worked well in practice, so we set \texttt{z\_proj\_linear} to true in our experiments. To enhance the model's robustness and prevent overfitting, we incorporated an optional dropout mechanism in the projection layers, though we found that for our primary experiments, setting the dropout rate to 0 yielded optimal results.

\paragraph{Contrastive Learning Approach} For training, we employed a contrastive learning objective where the model learned to identify acoustic boundaries by predicting future frames. We used a single-step prediction horizon (\texttt{pred\_steps=1}) with no offset (\texttt{pred\_offset=0}), which we found provided the most reliable boundary detection performance. The similarity between predicted and actual frames was measured using cosine similarity with a scaling coefficient of 1.0.

For each positive pair, we constructed a negative pair using a random permutation strategy. While our implementation supported both within-utterance and cross-utterance negative samples through the \texttt{batch\_shuffle} parameter, we found that using within-utterance negatives (setting \texttt{batch\_shuffle=false}) produced more consistent results, as it forced the model to learn fine-grained distinctions within the same acoustic context.

\paragraph{Training Details} The detector was trained on a speech dataset, with the primary experiments conducted using the Liberspeech dataset. We used the Adam optimizer with a learning rate of 0.0002, a batch size of 80, and trained for up to 200 epochs. Early stopping was employed based on validation performance to prevent overfitting. All training was conducted on NVIDIA V100 GPUs.

\paragraph{Boundary Detection Inference} During inference, the feature detector outputs similarity scores that undergo several post-processing steps to identify boundaries. The raw scores from prediction steps are combined and normalized using min-max normalization. Boundary detection is then performed using a peak detection algorithm, which identifies local maxima in the processed similarity scores. The key parameters for peak detection include a prominence threshold of 0.01, along with optional width and distance constraints that were automatically tuned during training.

\subsection{Details of Distinctive Codec}
The Distinctive Codec builds upon the SEANet-based encoder-decoder architecture used in SpeechTokenizer, extending it with our distinctive feature detection and variable-length segment processing capabilities. Here, we provide implementation details not covered in the main text.

\paragraph{Model Architecture} The encoder consists of a SEANetEncoder with 64 initial filters and a feature dimension of 1024. For our primary configuration yielding 9.5 tokens per second, we use strides of [8,5,4,2] (resulting in a total downsampling ratio of 320). For our higher frame rate configuration (15.7 tokens per second), we employ strides of [8,5,2,2], which produces a lower downsampling ratio and consequently more tokens per second. The encoder includes 2 bidirectional LSTM layers and a residual network with kernel size 3 and 1 residual layer per block. After encoding, the high-dimensional features (1024) are projected to a lower dimension (72) to make the subsequent distinctive feature processing more efficient.

The Distinctive Feature Encoder (DFE), implemented as the PerSegmentAutoEncoder in our code, compresses variable-length segments into fixed-length representations. The encoder component uses two convolutional layers with kernel size 3 and stride 1, followed by an adaptive average pooling operation to compress the temporal dimension to a single token. This architecture efficiently captures the salient information within each distinctive segment while maintaining a consistent output shape regardless of input segment length.

For quantization, we implemented Group-wise Scalar Quantization (GSQ) through our RefinedProjectionFSQ module, which divides the feature vector into multiple groups. Each group undergoes independent projection-based quantization to improve stability and representation quality. This approach was crucial for maintaining performance when operating at lower token rates.

The Distinctive Feature Decoder (DFD), also implemented within the PerSegmentAutoEncoder module, reconstructs the variable-length segments from the quantized representations. It uses nearest-neighbor interpolation to expand the fixed-length representations to their original temporal dimensions, followed by two convolutional layers with kernel size 3 and stride 1 to refine the expanded features. The decoder output is then projected back to the original high dimension (1024) before being processed by the SEANetDecoder, which mirrors the encoder structure to generate the final waveform.

\paragraph{Training Methodology} The Distinctive Codec was trained using a combination of reconstruction and perceptual losses:

\begin{itemize}
    \item \textbf{Time-Domain Reconstruction Loss}: We used L1 loss between the original and reconstructed waveforms, weighted by a factor of 500 to ensure accurate time-domain reconstruction.
    
    \item \textbf{Multi-resolution Mel-spectrogram Loss}: To capture perceptual qualities at different time scales, we employed a multi-resolution approach with four mel-spectrogram losses at different resolutions (using FFT sizes that vary by factors of 2). These losses combined L1 and L2 distances and were weighted at [45, 1, 1, 1] respectively, emphasizing the base resolution.
    
    \item \textbf{Adversarial Losses}: We employed multiple discriminators to improve the perceptual quality:
        \begin{itemize}
            \item Multi-Period Discriminators with periods [2, 3, 5, 7, 11]
            \item Multi-Scale Discriminators operating at different resolutions
            \item Multi-Scale STFT Discriminators analyzing the spectral characteristics
        \end{itemize}
    
\end{itemize}

\begin{table*}[t] 
\centering
\small
\def\arraystretch{1.3}
\setlength{\tabcolsep}{7pt}
\setlength{\abovetopsep}{0pt}
\setlength{\belowbottomsep}{0pt} 
\setlength{\aboverulesep}{0pt} 
\setlength{\belowrulesep}{0pt}
\caption{Performance comparison of frame-based and distinctive feature-based speech tokenization methods. Frame: frame rate (Hz); TKR: tokens per second~\cite{du2024funcodec}; BPS: bits per second~\cite{parker2024scaling}. RVQ: Residual Vector Quantizer~\cite{van2017neural}; FSQ: Finite Scalar Quantization~\cite{mentzerfinite}; GSQ: Group-wise Scalar Quantization (ours). Lower values are better for MEL Error, STFT, and WER; higher values are better for PESQ~\cite{pesq} and STOI~\cite{jensen2016algorithm}. All models use identical SEANet-based encoder-decoder architecture trained on LibriSpeech.}
\resizebox{\textwidth}{!}{%
\begin{tabular}{@{}c@{\hspace{10pt}}c@{\hspace{10pt}}c@{\hspace{10pt}}c@{\hspace{10pt}}c@{\hspace{10pt}}c@{\hspace{10pt}}c@{\hspace{10pt}}c@{\hspace{10pt}}c@{\hspace{10pt}}c@{}}
\toprule[1.0pt]
\multirow{2}{*}{\textbf{Segmentation}} & \multirow{2}{*}{\textbf{Quantization}} & \multirow{2}{*}{\textbf{Frame}} & \multirow{2}{*}{\textbf{TKR}} & \multirow{2}{*}{\textbf{BPS}} & \multicolumn{5}{c}{\textbf{Metrics}} \\
\cmidrule(lr){6-10}
 & & & & & MEL Error$\downarrow$ & STFT$\downarrow$ & PESQ$\uparrow$ & STOI$\uparrow$ & WER$\downarrow$ \\
\midrule
\multirow{5}{*}{\centering\textbf{Frame-based}} 
 & \centering RVQ & 10 & 10 & 100 & 0.4487 & 2.0183 & 1.2844 & 0.6695 & 0.9340 \\
 & \centering RVQ & 12.5 & 12.5 & 125 & 0.4290 & 1.9448 & 1.3245 & 0.6624 & 0.7701 \\
 & \centering RVQ & 20 & 20 & 200 & 0.3796 & 1.7401 & 1.4960 & 0.7225 & 0.6887 \\
 & \centering RVQ & 50 & 50 & 500 & 0.2342 & 0.5648 & 2.4496 & 0.8439 & 0.1698 \\
 & \centering FSQ & 20 & 20 & 320 & 0.1933 & 0.4699 & 2.5891 & 0.8615 & 0.1428 \\
\midrule
\multirow{4}{*}{\centering\textbf{\begin{tabular}[c]{@{}c@{}}Distinctive Feature\\-based (Ours)\end{tabular}}} 
 & \centering RVQ & 9.5 & 9.5 & 95  & 0.4481 & 2.0040 & 1.3312 & 0.6930 & 0.8523 \\
 & \centering FSQ & 9.5 & 9.5 & 152 & 0.4033 & 1.9042 & 1.4649 & 0.7049 & 0.6794 \\
 & \centering GSQ & 9.5 & 9.5 & 152 & 0.2857 & 1.3213 & 1.9147 & 0.7675 & 0.4265 \\
 & \centering GSQ & 15.7 & 15.7 & 251 & 0.2468 & 0.9072 & 2.3092 & 0.8203 & 0.2637 \\
\toprule[1.0pt]
\end{tabular}
}
\label{tab:speech_tokenization}
\end{table*}
\paragraph{Implementation Details} The model was trained for 20 epochs with a batch size of 9 using the Adam optimizer with learning rate 1e-4 and betas [0.9, 0.99]. Training was performed on LibriSpeech using 48000-sample segments (3 seconds at 16kHz) on 4 NVIDIA V100 GPUs. We used a cosine annealing learning rate schedule over the course of training. To ensure stable training, we found that initializing network weights with near-orthogonal initialization improved convergence. The model checkpoints were saved every 2,500 steps, with the final model selected based on the lowest validation mel-spectrogram error. The average token rate of our model is 9.5 tokens per second, with the actual rate varying based on the acoustic complexity of the input speech.

\subsection{Codec Evaluation Metrics}
We evaluate our Distinctive Codec using metrics for reconstruction quality, intelligibility, and encoding efficiency following previous works~\cite{zeghidour2021soundstream,defossezhigh}. For reconstruction, we measure mel-spectral error, STFT distance, and PESQ (Perceptual Evaluation of Speech Quality) scores~\cite{pesq}. For intelligibility, we measure Word Error Rate (WER) using the Whisper en-medium model~\cite{radford2023robust}, following SpeechTokenizer~\cite{zhang2024speechtokenizer}, and STOI (Short-Time Objective Intelligibility)~\cite{jensen2016algorithm} to quantify how accurately the speech content is preserved. We also track encoding efficiency through Token Ratio (TKR)~\cite{du2024funcodec}, representing tokens per second of 16 kHz audio, and Bits Per Second (BPS). The BPS calculation follows the approach in \cite{parker2024scaling} which will consider the vocabulary size of quantization method.

\subsection{Depression Detection Evaluation}
Evaluating the downstream performance of speech tokenizers for clinical applications presents fundamental methodological challenges. Training full-scale clinical assessment models requires substantial computational resources and carefully curated clinical data, making comprehensive comparison across multiple tokenization approaches prohibitively expensive. Furthermore, traditional codec evaluation metrics that focus on reconstruction quality and speech recognition fail to capture the preservation of subtle acoustic and temporal characteristics essential for mental health assessment, creating a significant evaluation gap between codec performance and clinical utility.

To enable systematic comparison of how different tokenization strategies preserve clinically relevant temporal information, we adopt the token projection evaluation framework proposed in~\cite{deshmukh2023pengi}. This methodology isolates the effects of tokenization by maintaining identical downstream architectures and training procedures across all evaluated methods, ensuring that performance differences directly reflect the quality of information preserved in discrete representations rather than variations in model design. We apply this comparative framework to the binary depression classification task using the DAIC-WOZ dataset~\cite{devault2014simsensei}, which requires capturing subtle prosodic variations, pause patterns, and speech rate fluctuations—the temporal dynamics that our distinctive feature-based approach is designed to preserve.

\paragraph{Evaluation Protocol.} Depression detection is evaluated on the DAIC-WOZ dataset, containing 107 training participants and 35 development set participants from clinical interviews. Following standard protocols, we extract participant speech segments (excluding interviewer turns) from each session and tokenize them using different codec methods under comparison. Depression labels are derived from PHQ-8 scores, with scores $\geq 10$ indicating clinical depression. For each tokenization method, we employ a Llama 3.1 8B model~\cite{dubey2024llama} with trainable projection module and classification components, trained using AdamW (lr=$5 \times 10^{-5}$, batch size 16) for 50 epochs with bf16 mixed precision. Participant speech from each session is processed as variable-length segments projected to 128 tokens. We report F1-score, UAR, and accuracy as evaluation metrics, with all experiments conducted three times using different random seeds and results averaged across runs.

\section{Experimental Results}
\subsection{Codec Results}

Table~\ref{tab:speech_tokenization} presents a systematic comparison between frame-based and distinctive feature-based tokenization approaches under identical architectural and training conditions. Our distinctive feature-based codec operates at average token rates of 9.5 Hz and 15.7 Hz, significantly lower than conventional fixed-rate processing while maintaining superior reconstruction quality.

The effectiveness of our distinctive feature-based approach is evident when comparing models at similar token rates. At 9.5 Hz, our method with RVQ quantization outperforms the frame-based baseline at 10 Hz across all metrics. Despite using a slightly lower token rate, our approach reduces MEL Error by 0.0006, decreases STFT distortion by 0.0143, improves perceptual quality (PESQ) by 0.0468, enhances intelligibility (STOI) by 0.0235, and reduces WER by 0.0817. These consistent improvements demonstrate that adaptive segmentation guided by acoustic boundaries preserves more acoustic information than arbitrary fixed-interval processing, even when using fewer tokens per second.

We observe further improvements when replacing RVQ with FSQ in our distinctive feature-based codec. Notably, frame-based approaches exhibit instability with FSQ at lower token rates (below 20 Hz), preventing direct comparison at our model's operating point of 9.5 Hz. This instability supports our hypothesis that fixed-rate segmentation at long intervals forces the quantizer to represent acoustically heterogeneous content within single frames, leading to training difficulties. The comparison between GSQ and FSQ at the same token rate (9.5 Hz) clearly demonstrates the effectiveness of our Group-wise Scalar Quantization approach. GSQ substantially outperforms FSQ across all metrics, reducing MEL error by 0.1176, decreasing STFT distortion by 0.5829, improving PESQ by 0.4498, increasing STOI by 0.0626, and lowering WER by 0.2529. This validates our hypothesis that decomposing high-dimensional quantization into multiple lower-dimensional sub-problems enhances stability and representation quality for variable-length segments generated by distinctive feature-based segmentation.

\subsection{Depression Detection Results}
\begin{table}[t]
\centering
\caption{Depression detection performance comparison on DAIC-WOZ development set. All models use identical downstream architecture (Llama 3.1 8B with projection module) and training protocol. Results are averaged over three random seeds.}
\label{tab:depression_detection}
\begin{tabular*}{\columnwidth}{@{\extracolsep{\fill}}lc@{}}
\toprule
\textbf{Tokenization Method} & \textbf{F1-Score} \\
\midrule
Frame-based (10Hz RVQ) & 0.471 \\
Distinctive Feature-based (9.5Hz RVQ) & 0.533 \\
Distinctive Feature-based (9.5Hz GSQ) & \textbf{0.636} \\
\bottomrule
\end{tabular*}
\end{table}
\begin{figure*}[t]
    \centering
    \begin{subfigure}[b]{0.48\textwidth}
        \centering
        \includegraphics[width=\textwidth]{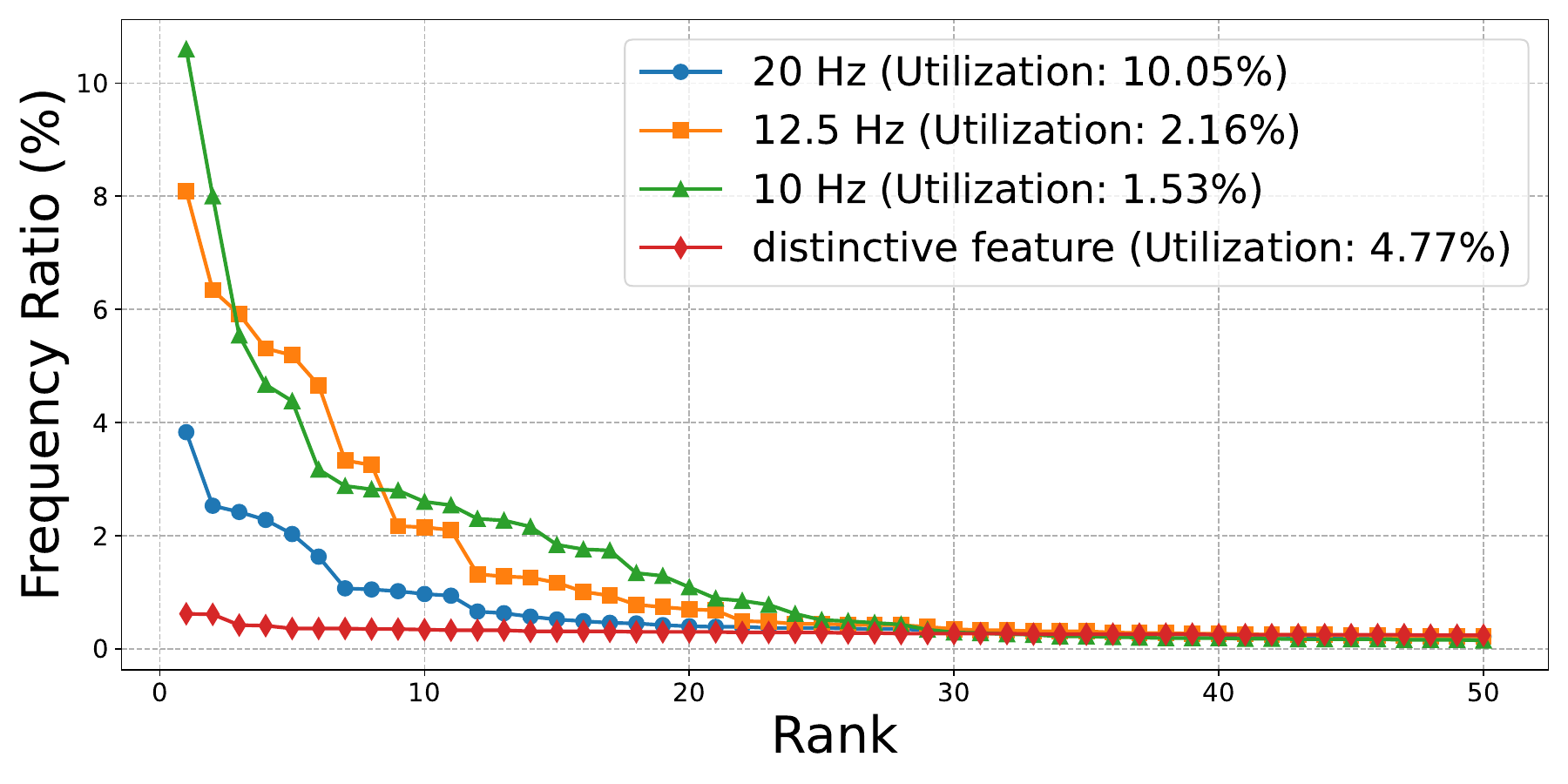}
        \caption{Top 50 Code Frequency Ratios with Their Overall Utilization Rates Shown in the Legend.}
        \label{fig:frequency}
    \end{subfigure}
    \hfill
    \begin{subfigure}[b]{0.48\textwidth}
        \centering
        \includegraphics[width=\textwidth]{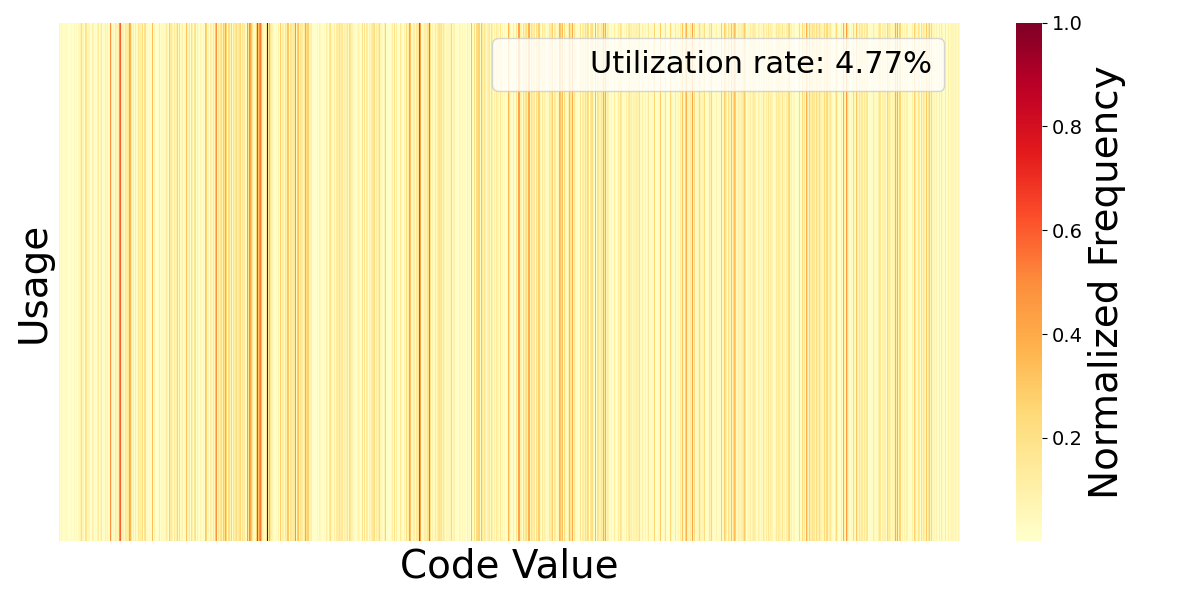}
        \caption{Visualization of Codebook Usage for Codec Using Distinctive Feature}
        \label{fig:usage}
    \end{subfigure}
    \caption{Codebook utilization comparison between frame-based and distinctive feature-based processing. Our approach achieves more balanced and efficient codebook usage (4.77\% utilization) compared to frame-based methods at similar frame rates (1.53\% at 10Hz).}
    \label{fig:codebook}
\end{figure*}
Table~\ref{tab:depression_detection} presents the depression detection performance comparison across tokenization approaches. Our distinctive feature-based codec with GSQ achieves an F1-score of 0.636, substantially outperforming both the frame-based baseline (F1=0.471, +35.0\% relative improvement) and our method with standard RVQ quantization (F1=0.533, +19.3\% relative improvement). These results provide direct evidence that timing-preserving tokenization captures clinically relevant temporal dynamics that are destroyed by fixed-rate processing.

The performance hierarchy reveals two complementary effects. First, comparing frame-based (F1=0.471) versus distinctive feature-based with RVQ (F1=0.533) isolates the contribution of adaptive segmentation: aligning token boundaries with acoustic transitions preserves the pause patterns, speech rate variations, and prosodic contours that serve as established biomarkers for depression. Second, the further gain from GSQ (F1=0.636) demonstrates that stable quantization of variable-length segments is crucial for preserving fine-grained temporal information at low token rates. Together, these findings validate that our approach—adaptive segmentation plus robust quantization—effectively encodes the temporal dynamics essential for clinical speech analysis.

\section{Ablation Study and Discussion}
\subsection{Analysis of Distinctive Feature Effectiveness}
\paragraph{Theoretical Derivation Hypothesis}
Consider an autoencoder that extracts a latent representation from input speech frames. For each segment of input frames $x \in \mathbb{R}^{D \times L}$, the encoder produces a latent vector $z = f(x) \in \mathbb{R}^d$. The autoencoder training objective and quantization distortion can be formulated as:
\begin{equation}
\min_{f, g} \; \mathbb{E}_{x}\left[ \| x - g(f(x)) \|^2 \right], \quad D = \mathbb{E}\left[\| z - Q(z) \|^2\right]
\end{equation}
where $g$ is the decoder and $Q$ is the quantizer that maps $z$ to a finite codebook. High-resolution quantization theory~\cite{gray1998quantization} approximates this distortion as:
\begin{equation}
D \approx G(d) \left( \int p(z)^{\frac{d}{d+2}} dz \right)^{\frac{d+2}{d}} 2^{-\frac{2R}{d}}
\end{equation}
where $G(d)$ is a dimension-dependent constant and $R$ is the bit rate. The integral term is critical, as it depends on the distribution of latent vectors.

Based on these formulations, we hypothesize that the effectiveness of distinctive features stems from their impact on the latent distribution $p(z)$. We posit that frame-based segmentation, which arbitrarily divides speech without regard to acoustic boundaries, may potentially force a single segment to capture multiple distinct speech states. This would cause the resulting latent vectors to represent a mixture of acoustic features, leading to a more diffuse distribution in the latent space. Under this hypothesis, our distinctive feature approach should yield more concentrated latent distributions by aligning segment boundaries with natural acoustic transitions. Specifically, when segments contain acoustically homogeneous content, the encoder can produce latent vectors that cluster more tightly around prototype representations of discrete speech units. Such a multimodal distribution would theoretically allow for more efficient quantization, as codebook entries could be optimally positioned to capture these distinct modes, thereby reducing the overall distortion $D$ for a given rate $R$.

\paragraph{Codebook Analysis}
Building on our theoretical analysis, we now examine the empirical evidence supporting our hypothesis through codebook utilization patterns. Since our theoretical framework suggests that distinctive features should allow for more efficient quantization by producing more concentrated latent distributions, analyzing codebook utilization provides a direct way to verify this effect in practice.

To investigate this hypothesis, we conducted experiments comparing codebook utilization across different processing approaches, with all models using FSQ for quantization. Figure \ref{fig:codebook} presents the utilization statistics for frame-based models operating at different frame rates (20Hz, 12.5Hz, and 10Hz) alongside our Distinctive Codec. The results reveal several important insights. As shown in Figure \ref{fig:frequency}, when frame rates decrease in conventional frame-based models, codebook utilization rates drop dramatically—from 10.05\% at 20Hz to merely 1.53\% at 10Hz. This declining utilization explains the instability we encountered when attempting to run Speech Tokenizer with FSQ at lower frame rates, as the quantizer struggles to effectively represent the diverse acoustic content when arbitrarily segmented at longer intervals.

In stark contrast, our Distinctive Codec achieves a substantially higher codebook utilization rate of 4.77\% despite operating at a comparable frame rate (9.5Hz) to the 10Hz frame-based model. This represents over three times better utilization of the quantization space. Moreover, the frequency distribution in Figure \ref{fig:frequency} shows that our approach exhibits a more balanced utilization pattern across codebook entries, indicating a more effective mapping of acoustic features to the discrete representation space. Conventional frame-based approaches show highly skewed distributions with a few dominant codes and many rarely-used entries, whereas our distinctive feature-based segmentation leads to a more uniform distribution. The visualization of the actual codebook usage in Figure \ref{fig:usage} further illustrates how our approach better leverages the available codebook capacity through perceptually-guided segmentation.

These empirical findings strongly support our theoretical hypothesis: by aligning segment boundaries with natural acoustic transitions, distinctive feature-based processing produces more coherent latent representations that can be more efficiently quantized. The improved codebook utilization directly translates to better reconstruction quality and speech intelligibility as demonstrated in our main experimental results, validating the fundamental advantage of our approach over uniform frame-based processing.

\subsection{Analysis of Group-wise Scalar Quantization Effectiveness}

\begin{table}[t]
\centering
\scriptsize
\def\arraystretch{1.25} 
\setlength{\tabcolsep}{3pt} 
\setlength{\abovetopsep}{0pt}
\setlength{\belowbottomsep}{0pt}
\setlength{\aboverulesep}{0pt}
\setlength{\belowrulesep}{0pt}
\caption{Comparison between FSQ and our proposed GSQ (many to many) at different reconstruction levels. Both methods are evaluated on the same Distinctive Codec architecture operating at 9.5 Hz.}
\label{tab:fsq_vs_gsq}
\resizebox{\columnwidth}{!}{
\begin{tabular}{@{}lc ccccc@{}}
\toprule[1.0pt]
\textbf{Method} & \textbf{Level} & \textbf{MEL}$\downarrow$ & \textbf{STFT}$\downarrow$ & \textbf{PESQ}$\uparrow$ & \textbf{STOI}$\uparrow$ & \textbf{WER}$\downarrow$ \\
\midrule
FSQ & 32 & 0.1768 & 0.5844 & 2.5467 & 0.8645 & 0.1301 \\
FSQ & 64 & 0.1407 & 0.4330 & 2.7909 & 0.8950 & 0.0732 \\
\midrule
GSQ & 24 & 0.2056 & 0.6963 & 2.3578 & 0.8474 & 0.1794 \\
GSQ & 32 & 0.1872 & 0.6207 & 2.4732 & 0.8597 & 0.1433 \\
GSQ & 64 & 0.1456 & 0.4423 & 2.7848 & 0.8889 & 0.0749 \\
\toprule[1.0pt]
\end{tabular}
}
\end{table}
\begin{figure*}[t]
    \centering
    \begin{subfigure}[b]{0.45\textwidth}
        \centering
        \includegraphics[width=\textwidth]{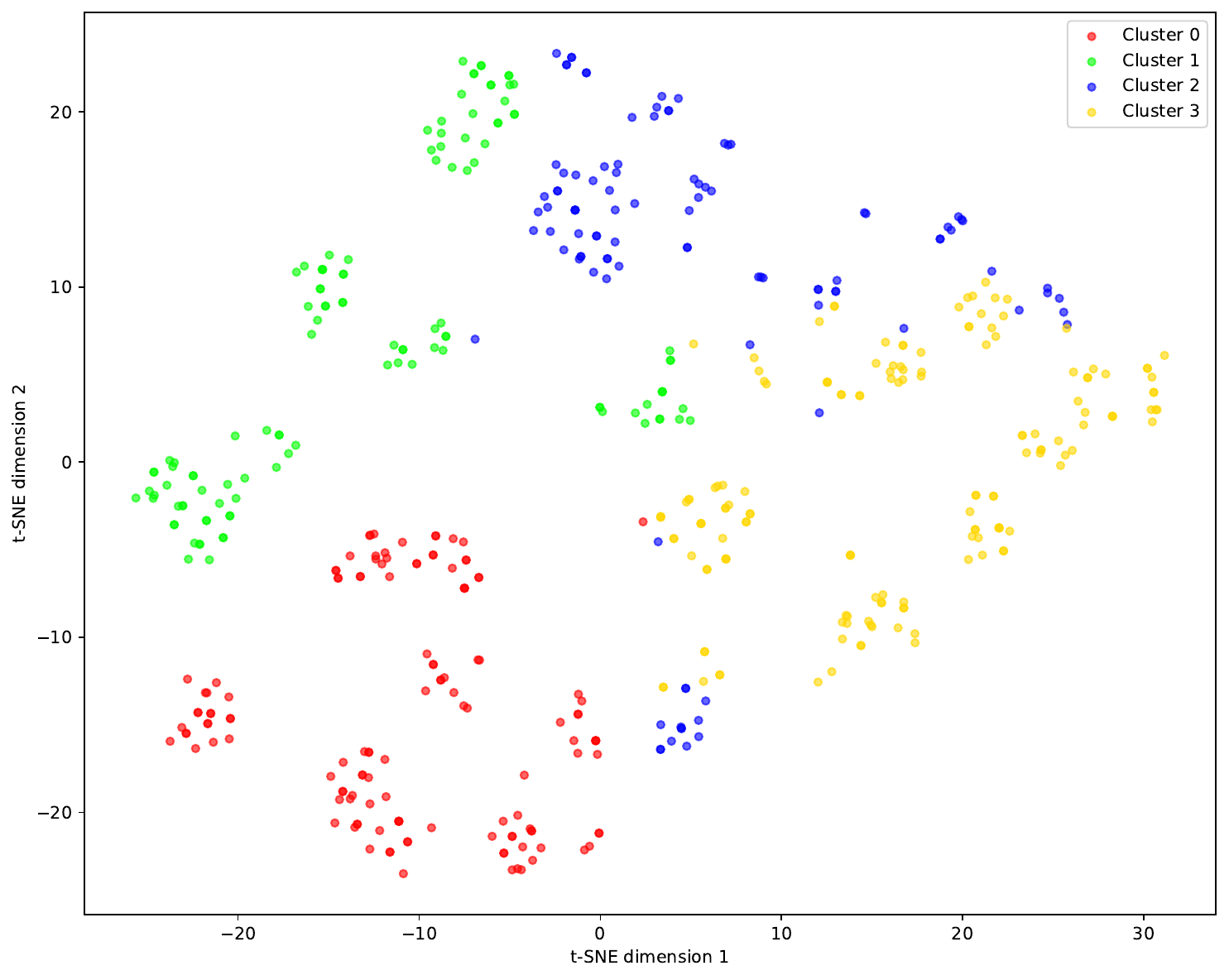}
        \caption{FSQ latent space clustering}
        \label{fig:fsq_cluster}
    \end{subfigure}
    \hfill
    \begin{subfigure}[b]{0.45\textwidth}
        \centering
        \includegraphics[width=\textwidth]{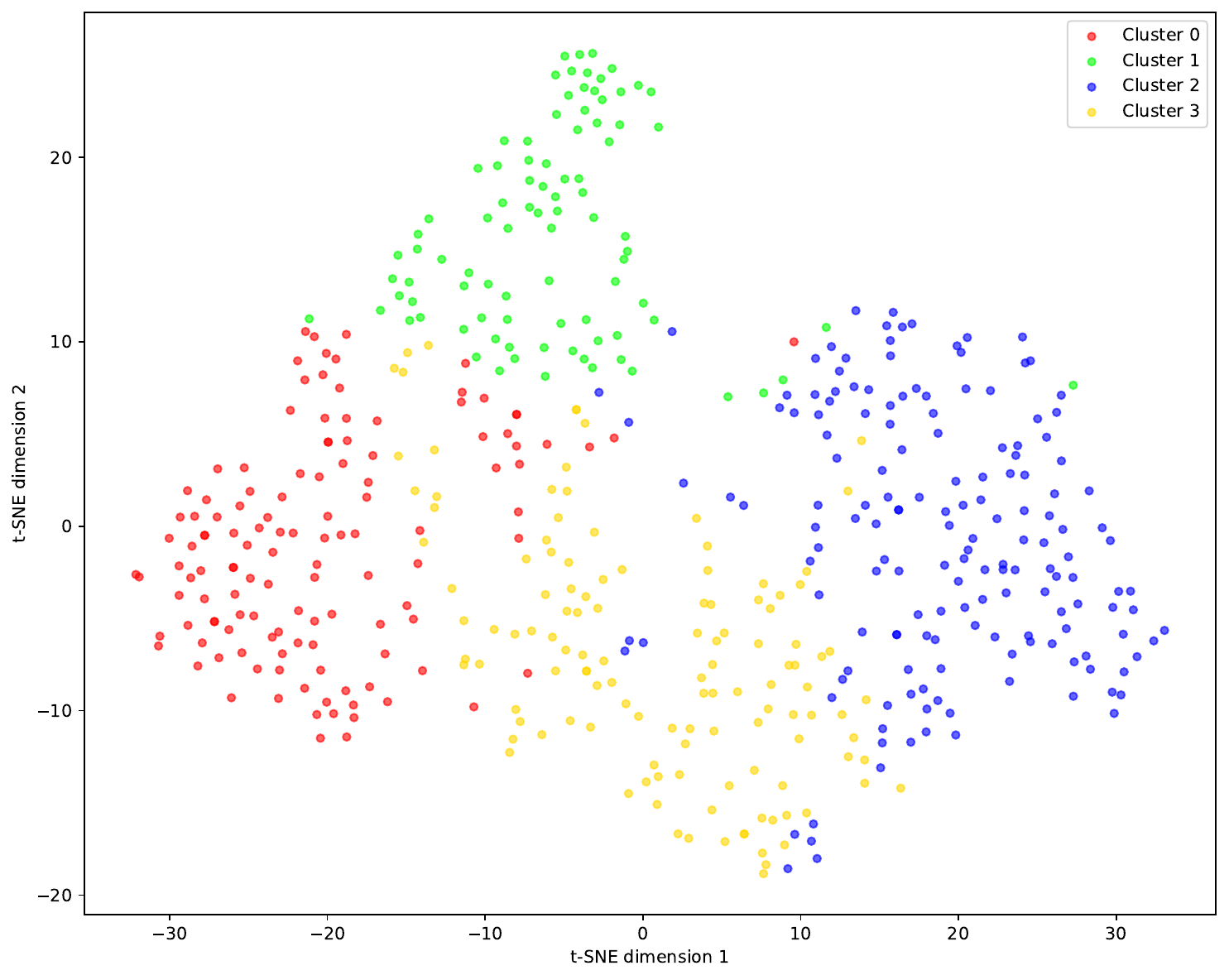}
        \caption{GSQ latent space clustering}
        \label{fig:subfsq_cluster}
    \end{subfigure}
   \caption{t-SNE visualization~\cite{van2008visualizing} of latent space clusters. (a) Standard FSQ shows overlapping clusters. (b) GSQ produces more distinct clusters with clearer boundaries.}
   \vspace{-18pt}
    \label{fig:cluster}
\end{figure*}
\paragraph{Comparative Analysis of Quantization Methods}
Building on our theoretical analysis of distinctive features, we next investigate the effectiveness of our quantization strategy when applied to segments of varying complexity. Table \ref{tab:fsq_vs_gsq} compares our GSQ approach with standard FSQ across different reconstruction levels. While our primary results focused on compressing variable-length segments into single tokens, this analysis examines how effectively both methods preserve information at different dimensionalities. The reconstruction level represents the latent space dimension to which each segment is compressed before quantization. Our GSQ approach offers a significant advantage by dramatically reducing the effective dictionary size required for high-quality representations. Whereas direct application of FSQ to high-dimensional features would demand an extremely large dictionary, our group-wise strategy effectively partitions the feature space into multiple specialized quantization subproblems. This approach not only enhances computational efficiency but also improves representation stability for distinctive feature-based segments. Despite using a much smaller effective dictionary, GSQ achieves performance comparable to FSQ across all metrics, particularly at higher reconstruction levels. 
\paragraph{Analysis of GSQ's Effectiveness Through Latent Space Geometry}
We hypothesize that our GSQ approach outperforms standard quantization methods due to two key information-theoretic advantages. First, by projecting high-dimensional features onto specialized lower-dimensional subspaces before quantization, GSQ maximizes mutual information between input and quantized output. The learned projection matrices effectively discard redundant information while preserving essential acoustic patterns, resulting in a lower KL divergence between original and quantized distributions. Second, unlike standard FSQ which quantizes each dimension independently, GSQ's group-wise approach exploits statistical dependencies in the data, effectively aligning quantization axes with the signal's intrinsic structure. This results in quantization cells that better fit the data distribution, reducing overall distortion.

Our information-theoretic analysis suggests that if GSQ preserves more structural information, this should be reflected in the geometric organization of the latent space~\cite{saxe2019information,goldfeld2020information}. Specifically, a quantization method that maintains higher mutual information and lower distortion should produce more coherent and well-separated clusters when visualized. To test this hypothesis, we employed t-SNE visualization~\cite{van2008visualizing}, which maps high-dimensional data to two dimensions while preserving local neighborhood relationships, making it ideal for assessing how well structural information is maintained after quantization. Figure \ref{fig:cluster} shows striking differences between standard FSQ (Figure \ref{fig:fsq_cluster}) and our GSQ approach (Figure \ref{fig:subfsq_cluster}). GSQ produces significantly more coherent clusters with clearer boundaries, indicating that GSQ preserves class structure and separability in the latent space. These results directly validate our hypothesis: GSQ's decomposition strategy prevents information mixing across feature groups, enabling it to preserves more meaningful information within severe token constraints. The improved geometric organization of GSQ's latent space explains its superior reconstruction quality across all our evaluation metrics.

\subsection{Generalization to Out-of-Domain and Code-Switched Speech}
\label{cs data}
To evaluate the generalization capability of our model beyond English and the LibriSpeech training domain, we conduct a zero-shot inference experiment on the SEAME dataset~\cite{lyu2010seame}, a 200-hour spontaneous Mandarin-English code-switching corpus widely used in multilingual speech research. SEAME poses three primary challenges: (i) Conversational spontaneity: frequent disfluencies (hesitations, overlaps, false starts) and colloquial prosody; (ii) Telephony distortions and environmental noise: low-bitrate channel artifacts and background sounds typical of mobile or landline settings; (iii) Cross-lingual code-switching: rapid alternation between Mandarin and English within utterances.

Without any fine-tuning, we directly apply our Distinctive Codec model trained on LibriSpeech to a subset of 50 utterances randomly selected from SEAME. The evaluation focuses on PESQ, which serves as the primary perceptual quality metric and correlates strongly with both intelligibility and signal fidelity.

Under a low token rate setting (9.5 Hz), our model achieves a PESQ score of 1.4214, significantly outperforming SpeechTokenizer's PESQ score of 1.0695 under the same configuration. This substantial improvement demonstrates that the proposed distinctive feature-based tokenization not only preserves critical acoustic cues in the English domain but also exhibits robust transferability to unseen languages and mixed-lingual acoustic conditions.

These results provide empirical evidence that Distinctive Codec retains perceptually significant information in cross-lingual and code-switched scenarios, supporting its potential as a universal speech tokenizer across diverse linguistic domains.

\subsection{Comparison with WaveTokenizer}
\label{WaveTokenizer}
\begin{table}[ht]
\centering
\scriptsize
\def\arraystretch{1.25} 
\setlength{\tabcolsep}{3pt} 
\setlength{\abovetopsep}{0pt}
\setlength{\belowbottomsep}{0pt}
\setlength{\aboverulesep}{0pt}
\setlength{\belowrulesep}{0pt}
\caption{Performance comparison between WaveTokenizer and our Distinctive Codec at similar token rates. Lower values are better for MEL Error, STFT, and WER; higher values are better for PESQ and STOI.}
\label{tab:wavetokenizer_comparison}
\resizebox{\columnwidth}{!}{
\begin{tabular}{@{}lc ccccc@{}}
\toprule[1.0pt]
\multirow{2}{*}{\textbf{Model}} & \multirow{2}{*}{\shortstack{\textbf{Token}\\\textbf{Rate}}} & \multicolumn{5}{c}{\textbf{Metrics}} \\
\cmidrule(lr){3-7}
 & & \textbf{MEL}$\downarrow$ & \textbf{STFT}$\downarrow$ & \textbf{PESQ}$\uparrow$ & \textbf{STOI}$\uparrow$ & \textbf{WER}$\downarrow$ \\
\midrule
WaveTokenizer & 10 Hz & 0.3139 & 1.0459 & 1.8333 & 0.7449 & 0.5535 \\
Distinctive Codec (GSQ) & 9.5 Hz & 0.2006 & 0.6021 & 2.1901 & 0.8114 & 0.3061 \\
\toprule[1.0pt]
\end{tabular}
}
\end{table}
To further validate the effectiveness of our distinctive feature-based approach, we conducted additional experiments comparing our method with WaveTokenizer~\cite{ji2024wavtokenizer}, another state-of-the-art neural speech codec. While our main results (Table \ref{tab:speech_tokenization}) demonstrate competitive performance against SpeechTokenizer, this additional comparison provides broader context for our approach within the current landscape of speech tokenization methods.

For a fair comparison, we trained WaveTokenizer on the LibriSpeech dataset with a token rate of 10 Hz, matching the operating conditions of our Distinctive Codec (9.5 Hz). The results are presented in Table \ref{tab:wavetokenizer_comparison}.

Despite operating at a slightly lower token rate, our Distinctive Codec with GSQ significantly outperforms WaveTokenizer across all evaluation metrics. Specifically, our approach reduces MEL Error by 36.1\%, STFT distortion by 42.4\%, and WER by 44.7\%, while improving PESQ by 19.5\% and STOI by 8.9\%. These substantial improvements further validate the effectiveness of our distinctive feature-based approach and Group-wise Scalar Quantization method.

\subsection{Investigate the impact of semantic distillation}

\begin{table}[ht]
\centering
\scriptsize
\def\arraystretch{1.25} 
\setlength{\tabcolsep}{3pt} 
\caption{Impact of Semantic Distillation (SD) on speech tokenization models. SD refers to the process of guiding the first RVQ layer with HuBERT representations. Lower values are better for MEL Error, STFT, and WER; higher values are better for PESQ and STOI.}
\label{tab:semantic_distillation}
\resizebox{\columnwidth}{!}{
\begin{tabular}{@{}l c ccccc@{}}
\toprule[1.0pt]
\multirow{2}{*}{\textbf{Model}} & \multirow{2}{*}{\shortstack{\textbf{Frame}\\\textbf{Rate}}} & \multicolumn{5}{c}{\textbf{Metrics}} \\
\cmidrule(lr){3-7}
& & \textbf{MEL}$\downarrow$ & \textbf{STFT}$\downarrow$ & \textbf{PESQ}$\uparrow$ & \textbf{STOI}$\uparrow$ & \textbf{WER}$\downarrow$ \\
\midrule
Distinctive Codec & 9.5 & 0.286 & 1.321 & 1.915 & 0.768 & 0.427 \\
Distinctive Codec + SD & 9.5 & 0.337 & 1.660 & 1.721 & 0.758 & \textbf{0.359} \\
\midrule
Speech Tokenizer & 10 & 0.314 & 1.046 & 1.833 & 0.745 & 0.554 \\
Speech Tokenizer + SD & 10 & 0.561 & 2.242 & 1.070 & 0.606 & 0.999 \\
\midrule
Speech Tokenizer & 50 & 0.234 & 0.565 & 2.450 & 0.844 & 0.170 \\
Speech Tokenizer + SD & 50 & 0.295 & 0.750 & 2.220 & 0.811 & \textbf{0.110} \\
\bottomrule[1.0pt]
\end{tabular}
}
\end{table}

To investigate the impact of semantic distillation on speech tokenization models, we conducted experiments comparing performance with and without distillation across different frame rates. Table~\ref{tab:semantic_distillation} presents these results, revealing several important insights not previously reported in the original SpeechTokenizer work.

Our findings demonstrate a consistent pattern: when semantic distillation is applied, reconstruction quality metrics (MEL Error, STFT, PESQ, STOI) tend to decrease, while speech content preservation measured by WER improves. This trade-off is evident in both our Distinctive Codec at 9.5 Hz and SpeechTokenizer at 50 Hz, where the addition of semantic distillation increases reconstruction errors but significantly reduces word error rates.

Notably, the comparison between SpeechTokenizer at 10 Hz and Distinctive Codec at 9.5 Hz highlights the superior stability of our approach at lower frame rates. While both models experience changes when semantic distillation is applied, SpeechTokenizer at 10 Hz shows dramatic degradation across all metrics. In contrast, our Distinctive Codec maintains relatively stable performance with more moderate reconstruction quality decreases and substantial WER improvements.

\section{Conclusion}
In this paper, we investigated the effectiveness of distinctive features for speech representation in depression detection, demonstrating that adaptive segmentation aligned with acoustic boundaries preserves critical temporal dynamics. Our experiments address the fundamental limitation of fixed-rate processing that destroys timing information essential for clinical applications. Results show that distinctive feature-based tokenization produces more coherent latent representations with improved codebook utilization, while our Group-wise Scalar Quantization strategy enables stable quantization at low token rates. The substantial performance gains in depression detection (+35.0\% relative improvement) validate that preserving temporal structure is crucial for capturing clinically relevant biomarkers such as pause patterns and speech rate variations. This work establishes distinctive features as a promising direction for neural speech codecs in temporally sensitive applications, with potential benefits for clinical assessment systems and other domains requiring faithful preservation of timing dynamics.

\section*{Acknowledgement}
This work was supported by the Australian Research Council Discovery Project DP230101184.

\bibliographystyle{IEEEtran}
\bibliography{reference}

\end{document}